# Synthesis of Ni doped ZnO nanostructures by low temperature wet chemical method and their enhanced field emission properties

Amit Kumar Rana[1], Prashant Bankar[2, 3], Yogendra Kumar[1], Mahendra A. More[3], Dattatray J. Late[2], Parasharam M. Shirage[1, *]

[1]Department of Physics & Metallurgical Engineering and Material Science, Indian Institute of Technology Indore, Simrol Campus, Khandwa Road, Indore-453552. India.

[2]Physical and Materials Chemistry Division, CSIR-National Chemical Laboratory, Dr. Homi Bhabha Road, Pashan, Pune 411008. India.

[3]Department of Physics, Savitribai Phule Pune University, Pune 411007. India.

[*]Author for correspondence: *pmshirage@iiti.ac.in, paras.shirage@gmail.com*

Email: - *datta099@gmail.com, dj.late@ncl.res.in; mam@physics.unipune.ac.in; aramitrana4@gmail.com;*

**Abstract:** In this study, we report the enhancement in field emission (FE) properties of ZnO nanostructure by Ni doping at the base pressure of ~1 x $10^{-8}$ mbar, which were grown by the simple wet chemical process. ZnO nanostructure shows single crystalline wurtzite structure up to Ni 10% doping. FESEM represents change in the nanostructure morphology from thick nanoneedles to nanoflakes via thin nanorods with increase in Ni doped ZnO. The turn-on field required to draw a field emission (FE) current density at 1µA/cm$^2$ is found to be 2.5, 2.3, 1.8 and 1.7 V/µm for ZnO(Ni0%), ZnO(Ni5%), ZnO(Ni7.5%) and ZnO(Ni10%), respectively. The highest current density of ~870 µA/cm$^2$ is achievable, which drawn at an applied field of 3.1 V/µm for Ni 10% doped ZnO. The long term operational current stability recorded at the preset values of 5 µA for 3 hr. duration and is found to be very good. The pragmatic results exhibit that the Ni doped ZnO based field emitter can open up many opportunities for their potential application as an electron source in flat panel display, transmission electron microscope, and X-ray generation. Thus, the simple wet chemical at low temperature (~80 °C) synthesis approach and robust nature of ZnO nanostructure field emitter can provide prospects for the future development of cost effective electron sources.

# 1. Introduction

Zinc oxide is one of the most striking semiconductors in the category of 1D and 2D nanomaterial. It is due to its remarkable and multifunctional properties such as direct bandgap of 3.37eV, high exciton binding energy 60 $meV$[1,2]. ZnO also shows numerous morphology such as nano-combs[3,4], nanotubes[5], nano-springs[6], nanorods[7], nanorings[8] nanowires[9] and nanoflower[10], due to its significant physical and chemical properties, it's a potential candidate for various applications, such as photo-electronics devices[11,12], nano-sensors[13-17], nano-generators[18], electronics devices[19,20], dye-sensitized solar cells[21], spintronics[22,23], metal-ions detection[24], photo-catalytic[25], *etc*. The last few decades ZnO has shown one of the most promising materials for FE device. It is because of high thermal stability, low electron affinity, as well as oxidation resistance in harsh environment[26,27]. There are many reports on enhancement of FE from ZnO *e.g.* Farid *et al.* have observed enhanced FE properties of Cu doped ZnO nano-composite films synthesized by electrochemically[28] and Yu-Cheng Chang, grown a Ni-doped ZnO nanotower arrays on a silicon substrate using thermal evaporation method (*T = 1100 ºC*) and observed enhanced optical and field emission properties[29]. Similarly, Xing *et al*. have reported the growth of ultrathin single-crystal ZnO nanobelts by using an Ag catalyzed vapor transport method and FE property[30]. The key issues are the operating voltage range and the emission current stability. As the turn-on or threshold field values are dependent on the morphology (shape and size) as well as its intrinsic physical properties such as work function and electrical conductivity of the emitter. The approach towards improving the FE properties of semiconductors is either to tailor the geometry of the emitter or by modifying its electronic properties. The first approach has limitations on the reduction of the size and shape of the nanostructures. Although the synthesis of an array of well-spaced anisotropic nanostructures possessing a very fine apex radius (typically, 20 nm) is feasible, the mechanical sturdiness of such an emitter is questionable. It is predictable that the mechanical stress induced due to the presence of an intense electrostatic field may result in negative effects, such as bending/fracture of the emitters. Similarly, for the second approach, modified electronic properties by doping/mixing with suitable elements are very critical and sometimes the chemical reactivity of such structures may lead to a decline of their emission performance. Therefore, a suitable combination of both approaches is considered to be a promising method for improving the FE performance, which has been well attempted by several researchers. Here we report the synthesis of Ni doped ZnO by low temperature wet chemical methods with the enhancement in FE behavior. We found after Ni doping turn on field decrease up to ~1.7 V$\mu$m$^{-1}$ and increase in emission current density of 872$\mu$A-cm$^{-2}$ under the field 3.1 V$\mu$m$^{-1}$. This significant enhancement in FE property is a due change in morphology as well as increase the number of electron in the conduction band.

## 2. Experimental details

Pure and Ni doped ZnO nanostructures were synthesized by low temperature (synthesis temperature 80°C) wet chemical method. Analytical grade zinc acetate dehydrates, nickel nitrate hexa-hydrate were used as the raw materials. 100 mM solution of zinc nitrate was prepared then to dope Ni, used appropriated amount of nickel nitrate was added in zinc nitrate, which were dissolved in 100 ml double distilled water and then stirred. After constant stirring for 30min, aqueous ammonia solution was added dropwise continuously to the solution till the pH approaches ~11. Finally, a clear solution was obtained in the case of pure ZnO and sea-blue solutions for Ni doped solutions. Then the solutions were kept at 80 °C for 2h, and then the precipitates were separated, dry for overnight at room temperature and finally dried sample is annealed at 150 °C (for 2 h), the detailed synthesis already reported elsewhere[31]. The samples with different Ni doping, 0%, 5%, 7.5% and 10% were named as ZNi0, ZNi5, ZNi7.5 and ZNi10, respectively. These the final composition of the materials after analysis by EDX. The phase purity and morphology of ZnO nanostructure were investigated by x-ray diffraction (Bruker D8 Advance) and field emission scanning electron microscope (Supra 55 Zeiss). The field emission investigations were carried out in a planar diode configuration in all metal ultrahigh vacuum (UHV) chambers which were evacuated to a base pressure of ~1x10$^{-8}$ mbar. A typical 'diode' configuration consists of a phosphor coated indium tin oxide semitransparent glass disc (a circular disc having diameter ~50 mm) which acts as an anode. The Ni doped ZnO nanostructure sprinkled onto a piece of UHV compatible conducting carbon tape pasted on a copper rod holder (diameter ~5 mm) served as the cathode. The emission current was measured by a Keithley electrometer (6514) by sweeping dc voltage applied to the cathode with a step of 40 V (0-40 kV, Spellman, U.S.). The stability of field emission current was investigated using a computer controlled data acquisition system with a sampling interval of 10 seconds. Special care was taken to avoid any leakage current using shielded cables and proper grounding. The field emission images were recorded using a digital camera (Canon SX150IS).

## *Ni-ZnO nanostructure growth mechanism:*

The possible growth mechanism of the nanomaterials can be described on basis of chemical reactions and nucleation. The perceptible reaction mechanisms are proposed for achieving the doped material oxides, which are proposed below:

$$Zn(NO_3)_2 + 2H_2O \xrightleftharpoons{NH_3,\ 80\ °C} Zn(OH)_2 + 2HNO_3 \quad \text{----------- (1)}$$

$$Ni(NO_3)_2 + 2H_2O \xrightleftharpoons{NH_3,\ 80\ °C} Ni(OH)_2 + 2HNO_3 \quad \text{----------- (2)}$$

$$Zn(OH)_2 + Ni(OH)_2 \rightarrow Ni-ZnO \downarrow + 3H_2O \uparrow \quad \text{----------- (3)}$$

$$Zn(OH)_2 \leftrightarrow Zn^{2+} + 2OH^- \quad \text{----------- (4)}$$

$$Zn^{2+} + 4NH_3 \leftrightarrow Zn(NH_3)_4^{2+} \quad \text{----------- (5)}$$

$$Zn^{2+} + 4OH^- \leftrightarrow Zn(OH)_4^{2-} \quad \text{----------- (6)}$$

$$Zn(OH)_4^{2-} \leftrightarrow ZnO + H_2O + 2OH^- \quad \text{----------- (7)}$$

$$Zn^{2+} + 2OH^- \leftrightarrow Zn(OH)_2 \leftrightarrow ZnO + 2H_2O \quad \text{----------- (8)}$$

$$Zn(NH_3)_4^{2+} + 2OH^- \leftrightarrow ZnO + 4NH_3 + 2H_2O \quad \text{----------- (9)}$$

These are the possible reactions which are responsible for the growth of ZnO nanostructure[16, 32]. The growth mechanism of ZnO nano-structures (nano-rods, nanoflakes, nano-combs, nano-springs, nano-belts, doughnut-shaped particles etc.) is controversial and still not very clear. The morphology of ZnO nanostructure is not only dependent on the preparation technique but also other external conditions such as reaction temperature, pH value, reaction time, solution concentration, doping, *etc*[5, 7]. The most plausible mechanism known for ZnO nano structure is either layer by layer along the axial direction or by screw dislocation, we already discuss both mechanisms in our previous reports[5, 7, 32]. ZnO exhibit several crystal planes with different polarity (one basal polar oxygen face (000$\bar{1}$), one polar zinc face (0001), six nonpolar (01$\bar{1}$1) faces and six symmetric non-polar faces (01$\bar{1}$0) parallel to c-axis). Wen-Jun Li *et al*.[33] shows different polar face have different growth velocity i.e. v (001) > v (01$\bar{1}$) > v (010) > v (011) > v (00$\bar{1}$). So, the growth along (002) is faster than other polar faces and forms hexagonal rods like morphology. When the Ni doped into the ZnO matrix, it will slow down growth direction along (002) and increase its thickness, so the average diameter of nanorods is slightly increased. While with higher Ni doping (10%), rods morphology starts altering into flakes with change in orientation along (100) and (101) direction. Due to high concentration of Ni ions in the solution, the nucleation in ZnO structure become easier to the lower activation energy barrier of heterogeneous nucleation[24]. This assists to grow ZnO in a different direction as a result to form flakes like morphology.

## 3. Results and discussion

Fig.1 (*a*) shows the XRD pattern of the pure and doped sample. All the samples are phase pure and exhibit a hexagonal phase of wurtzite type of ZnO, with preferred *c*-axis growth direction. There is no impurity peak found related to Ni phase, which indicates successfully incorporation of Ni in ZnO structure, it is due comparable ionic radii of $Ni^{2+}$ (0.69Å) and $Zn^{2+}$ (0.74Å). The most intense peak in ZnO is *(002)*; this is due to rods or needles like morphology of ZnO nanostructures[5, 29]. Furthermore with increasing the Ni concentration there is a decrease in intensity of *(002)* peak and increase in *(100)* and *(101)* peaks, which is due to change in orientation of nanostructure or morphology of ZnO doping, which is pointed out in previous studies[1, 4]. Fig.1 (*b*) represents the shift in *(002)* peak towards the lower two thetas up to 7.5% doping while Ni 10% doping displays an increase in two theta value. The reason may be different doping concentration from dissimilar conditions as well as a lot of distortions in the host ZnO lattices. This creates a lattice relaxation or compression in the host matrix, because of the vacancy or interstitial defect present in the host matrix. The similar trend of change in *2θ* value after Ni doping was reported by Yu *et al.*[34] and Tong *et al.*[35] From the interplanar spacing obtained XRD used to estimate the lattice parameters. The (002) plane XRD peak was used for the grain size calculation and the results are shown in table I.

Fig. 2 displays SEM images of pure and Ni doped ZnO nanostructures (magnification 50 KX). In a wet chemical method, the structure and morphology ZnO nanostructure depend upon external parameters such as reaction temperature, reaction time, *pH* value , solution concentration, and dopant concentration, *etc.* .These parameter plays crucial role in a physical and chemical properties of ZnO, which we already discuss in the previous report[4]. A typical SEM image of pure ZnO (Fig. 2(*a*)) (low magnification ( 5 KX) images are shown in *fig. S1*) shows the formation of nanoneedles in the form of flower like structure. The average diameter (from the middle) and the length were estimated to be ~200 nm and 1-2 μm, respectively. With Ni 5% and 7.5% doping nanoneedles converted into thick and top flat face nanoneedles with a slight increase in average diameter from 280 to 294 nm, respectively. It is also observed, cone tapered part of the nanoneedles cracked and starts developing into nanorods. Again with further increase in Ni concentration, nanoneedles is fully converted into nanorods and finally in Ni 10% doped shows decrease in average diameter of nanorods, less than ~100 nm with some transformation into nano-sheets in between the thin nanorods. The energy dispersive X-ray (EDX) spectra of all samples are shown in *fig*. S2 and percentage of Ni contain in the ZnO samples are shown in Table S1.

These 1D ZnO nanostructures show good field emitters because of their diverse nanostructure which provides promising aspect ratios and appropriate work functions. The FE properties can be altered by many factors, such as the curvature, uniformity, size, and density of emitter. The current density ($J$) versus applied electric field ($E$) characteristic of the pure and doped sample is shown in figure 3($a$). The values of turn-on field and threshold field required for drawing emission current densities of 1 and 10 $\mu A\ cm^{-2}$ are found to be 2.5 V$\mu m^{-1}$ and 2.8 V$\mu m^{-1}$, in the case of ZnNi0 and 1.7 V$\mu m^{-1}$ and 2 V$\mu m^{-1}$ for ZNi10 respectively. The emission current density is found to increase rapidly with increase in the applied electric field, and emission current density of 326 $\mu A\text{-}cm^{-2}$ and 872 $\mu A\text{-}cm^{-2}$ has been obtained at an applied field of 3.8 V-$\mu m^{-1}$ and 3.1 V $\mu m^{-1}$ for ZNi0 and ZNi10 respectively. FE characteristics of pure and doped ZnO sample are shown in table II.

The modified Fowler-Nordheim (F-N) equation in terms of the current density (J) and the applied electric field (E) where applied electric field E is defined as[36, 37].

$$J = \lambda_M a\phi^{-1} E^2 \beta^2 \exp(-\frac{b\phi^{\frac{3}{2}}}{\beta E} v_F)$$

Where E=V/d, where V is the applied voltage, and d is the separation between anode and cathode (~ 2 mm). Furthermore, the emission current density J is estimated as J=I/A, where, I is the emission current and A is the total area of the emitter, a and b are constants, typically $1.54\times10^{-10}$ (AV$^{-2}$eV) and $6.83\times10^3$ (V-eV$^{-3/2}\mu m^{-1}$) respectively, $\phi$ is the work function of the emitter material, $\lambda_M$ is the macroscopic pre-exponential correction factor, $v_F$ is value of the principal Schottky–Nordheim barrier function (a correction factor), and β is the field enhancement factor. In the present study, the FN plot is found to be nonlinear and such FN plots have been reported for many semiconductor nanomaterials (Fig.3 (b)).

The enhancement in FE performance after Ni doping in ZnO is may explain as follows: as we known, the turn-on or threshold field depends upon the overall geometry of emitter and also on the intrinsic electronic properties such as charge carrier concentration, work function, *etc*. ZnO rods with 5 and 7.5% Ni doping displays increment in the diameter relative to pure ZnO, whereas 10% Ni doped ZnO rods shows decrement. The overall morphology of ZNi5 and ZNi7.5 samples shows lower ariel density with quite separated rods which suggest less screening effect as compared to ZNi0. This decrement in screening results in deteriorating turn on voltage in both the cases. In ZNi10 sample, decreased diameter and presence of few nano-sheets which increases the number of emission centers, both these causes may results in a further decrement in turn on voltage. Another reason for decreased turn on voltage is the increased amount of Ni doping in ZnO which increases charge carrier concentration at the conduction band of ZnO as well as a shift in energy level that is proof by many experimental and theoretical studies[38]. Generally Ni isoelectronic with Zn, so it should not contribute to the acceptor or shallow donor into ZnO. Now the question is where the electrons come from? Katayama-Yoshida and Sato[39] show using ab initio electronic-structure calculation shows half metallic behavior exists in transition metal doped ZnO. In our sample, $Ni^{2+}$ partially substituted the $Zn^{2+}$ in ZnO and as we know $Zn^{2+}$ ion in ZnO structure located at the center of the tetrahedron surrounded by four oxygen atoms. Under the influence of tetrahedral crystal field of ZnO, d-states of Ni split into a higher triplet ($t_{2g}$) and lower doublet ($e_g$) state as shown in fig.4 (a). The higher $t_{2g}$ state hybridized with the p-orbital of the valance band and further split into $t_{bonding}$ and $t_{antibonding}$ (fig. 4(b)). The $t_{bonding}$ participate into the Ni-O bond and they are localized. But an antibonding state has higher energy and contains some itinerant electron and also the energy of the anti-bonding states lies very close to the conduction band. Hence there is the large probability that the electron from these antibonding states jumps (act as an impurity state) into the conduction band with a small increase in potential difference. Due to increases in Ni concentration into ZnO, more and more electron are promote to the conduction band, resulting increase in FE preformation as shown in schematic fig. 4(c & d) as compare to pure ZnO. In the present work this effect has been clearly observed from FE curve. Thus increased Ni concentration promotes a large amount of electrons in the conduction band of ZnO which ultimately results in decreasing turn on voltage for ZNi5, ZNi7.5 and ZNi10 compared to ZNi0.

Current stability curve of the pure and doped sample are also investigated at a preset value of 5 µA over a period of 3 *hrs.* (fig.5*)*.The successive current stability curve shows the no obvious degradation of current density. This is very important feature particularly for the practical application. The appearance of the "spikes" type fluctuations observed in the emission current may result from the adsorption/desorption and ion bombardment of residual gas molecules. The local work function of the emitter varies due to the adsorption/desorption gas molecule at the emitter surface and ion bombardment of residual gas molecules due to the presence of high electrostatic field resulting in fluctuations. Typical FE image is shown in inset *fig*.5. The image shows a number of tiny spots, corresponding to the emission from the most protruding edges of the emitters. These results exhibit the excellent emission stability of the Ni-doped ZnO, which make them highly valuable for practical applications as a field emitter. Furthermore, the Field emission behavior of other pure and doped ZnO nanostructure was compared with Ni doped ZnO data are shown in table III. Thus the overall FE performance of Ni-doped ZnO nanostructure prepared by low temperature wet chemical method in the present report, such as turn-on and threshold fields are all excellent than that of the previous report on pure and doped ZnO nanostructure.

## 4. Conclusion

In conclusion, Ni doped ZnO synthesis by simple and low temperature wet chemical method. XRD and SEM result confirms successfully incorporation of Ni ions in ZnO by variation in lattice constants as well as alteration in nanostructure morphology of ZnO nanoneedles, respectively. Ni doping in ZnO decreases the turn on the field (from 2.5 to 1.7 V$\mu m^{-1}$) and threshold field (from 2.8 to 2 V$\mu m^{-1}$) with maximum current density 872 $\mu A\ cm^{-2}$ at an applied field of 3.1 V $\mu m^{-1}$ for ZNi10. It is also long current stability for the period of 3 hrs. for the preset value of 5 $\mu A$. Therefore, these Ni-doped ZnO nanostructures are potential candidates for future application in nanoelectronics, particularly in front areas of flat panel displays and electron emitter devices.

## Acknowledgments

This work was supported by the Department of Science and Technology, India by awarding the prestigious 'Ramanujan Fellowship' (SR/S2/RJN-121/2012) to the PMS. PMS acknowledge the CSIR research grant No. 03(1349)/16/EMR-II. PMS is grateful to Prof. Pradeep Mathur, Director, IIT Indore, for encouraging the research work and providing the necessary facilities. Authors are thankful to SIC Indore for providing the research facilities like XRD and FESEM. P. K. Bankar acknowledges SPPU and DST for the financial support. Prof. M. A. More would like to thank the BCUD; of Savitribai Phule Pune University for the financial support provided for the field emission work under CNQS-UPE-UGC program activity. The research work was also supported by Department of Science and Technology (Government of India) under Ramanujan Fellowship to Dr. D. J. Late (Grant No. SR/S2/RJN-130/2012), CSIR-NCL-MLP project grant 028626, DST-SERB Fast-track Young scientist project Grant No. SB/FT/CS-116/2013, Broad of Research in Nuclear Sciences (BRNS) (Government of India) grants No. 34/14/20/2015 and the partial support by INUP IITB project sponsored by DeitY, MCIT, Government of India.

**Table**

**Table I.** Lattice parameters *a* and *c* of the Ni doped ZnO nanorods. Average grain sizes are calculated from the (002) reflection.

| Sample Name | a (Å) | c (Å) | c/a ratio | grain size(nm) |
|---|---|---|---|---|
| Pristine ZnO (ZNi0) | 3.249 | 5.201 | 1.6008 | 67.46 |
| ZnO with 5% Ni doped (ZNi5) | 3.253 | 5.206 | 1.6003 | 75.34 |
| ZnO with 7.5% Ni (ZNi7.5) | 3.254 | 5.210 | 1.6011 | 76.03 |
| ZnO with 10% Ni (ZNi10) | 3.246 | 5.197 | 1.6010 | 62.29 |

**Table II.** Comparison of field emission characteristics of pure and Ni doped ZnO nanostructures.

| Sr. No. | Sample Name | Turn on field (V µm$^{-1}$) at 1µA/cm$^2$ | Threshold Field (V µm$^{-1}$) at 10 µA/cm$^2$ | Maximum current density (µA/cm$^{-2}$) at V µm$^{-1}$ |
|---|---|---|---|---|
| 1 | ZNi0 | 2.5 | 2.8 | ~326 at ~3.8 |
| 2 | ZNi5 | 2.3 | 2.6 | ~156 at ~3.4 |
| 3 | ZNi7.5 | 1.8 | 2.2 | ~528 at ~3.4 |
| 4 | ZNi10 | 1.7 | 2 | ~872 at ~3.1 |

*Table III*. Comparison of field emission characteristics of the pure and doped ZnO nanostructure with the present work.

| *Field emitters* | *Synthesis route* | *Turn on Field ($V~\mu m^{-1}$)* | *Threshold Field ($V~\mu m^{-1}$)* | *References* |
| --- | --- | --- | --- | --- |
| ZnO nanoneedle | Vapor phase growth | 2.4 | 6.5 | 40 |
| ZnO nanowire | Vapor deposition method | 6.0 | 11.0 | 41 |
| Ga-doped ZnO | Vapor liquid solid process | 3.4 | 5.4 | 42 |
| In-doped ZnO | Chemical vapor deposition | 2.4 | 3.5 | 43 |
| N-doped ZnO | Solvothermal Synthesis | 2.9 | - | 44 |
| ZnO nanorods | Atomic layer deposition | 2.85 | - | 45 |
| ZnO nanowires/graphene | Hydrothermal method | 2.0 | - | 46 |
| Cu-doped ZnO quantum dot | Hydrothermal method | 4.47 | 8.9 | 47 |
| **Ni-doped ZnO** | **Wet chemical methods** | **1.7** | **2** | **This work** |

## Figure captions

***FIG.1***. *Shows (a) XRD of pure and Ni-doped ZnO nanostructure, (b) The relative shift in (002) with respect to Ni doping.*

***FIG.2***. *Shows the FESEM images (a) ZNi0 (b) ZNi5 (c) ZNi7.5 (d) ZNi10 (at 50 KX).*

***FIG.3***. *Shows Field emission of pure and Ni doped ZnO. (a) The applied electrical field as a function of emission current density. (b) F-N plot.*

***FIG.4.*** *Shows the schematic (a) Electronic structure of transition metal at a substitution site in a wurtzite structure. (b) Splitting of impurity state under the influence of crystal field of host ZnO. Schematic energy band diagram of (c) ZnO. (d) Ni doped ZnO ($E_g$: Energy gap, $E_f$: Fermi level, $E_c$: Energy of Conduction band, $E_v$: Energy of Valance band, $\phi$: work function).*

***FIG.5.*** *Shows typical field emission current stability recorded at 5 µA indicating stable field emission current of (a) pure ZnO, (b) Ni 5% doped ZnO, (c) Ni 7.5% doped ZnO, (d) Ni 10% doped ZnO (inset of fig. shows field emission pattern taken during the long term current stability measurements of the emitter).*

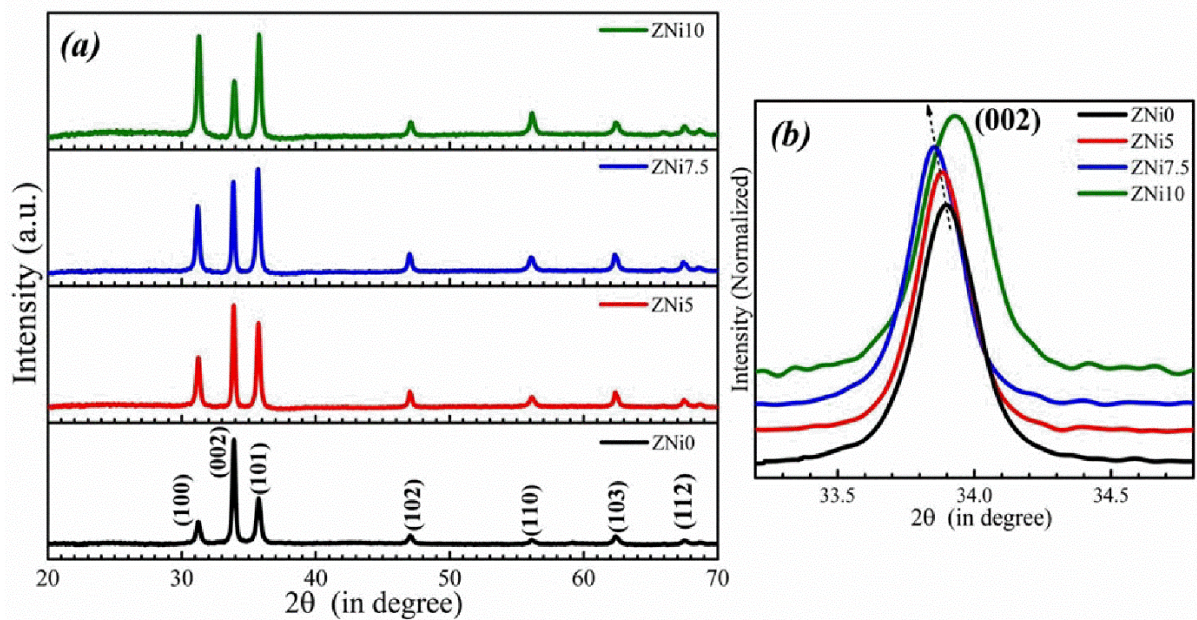

*FIG.1.Shows (a) XRD of pure and Ni-doped ZnO nanostructure, (b) the relative shift in (002) with respect to Ni doping.*

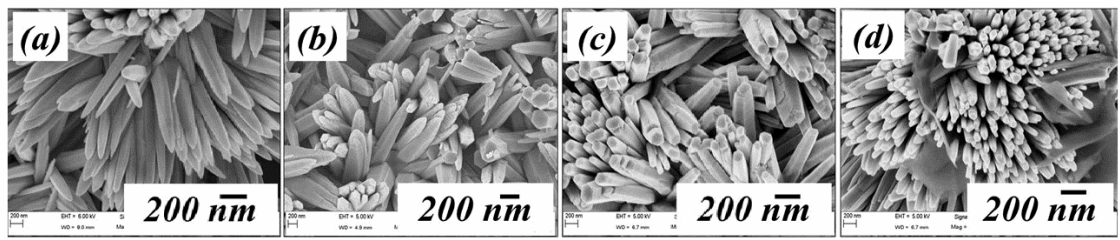

*FIG.2. Shows the FESEM images (a) ZNi0 (b) ZNi5 (c) ZNi7.5 (d) ZNi10 (at 50 K X).*

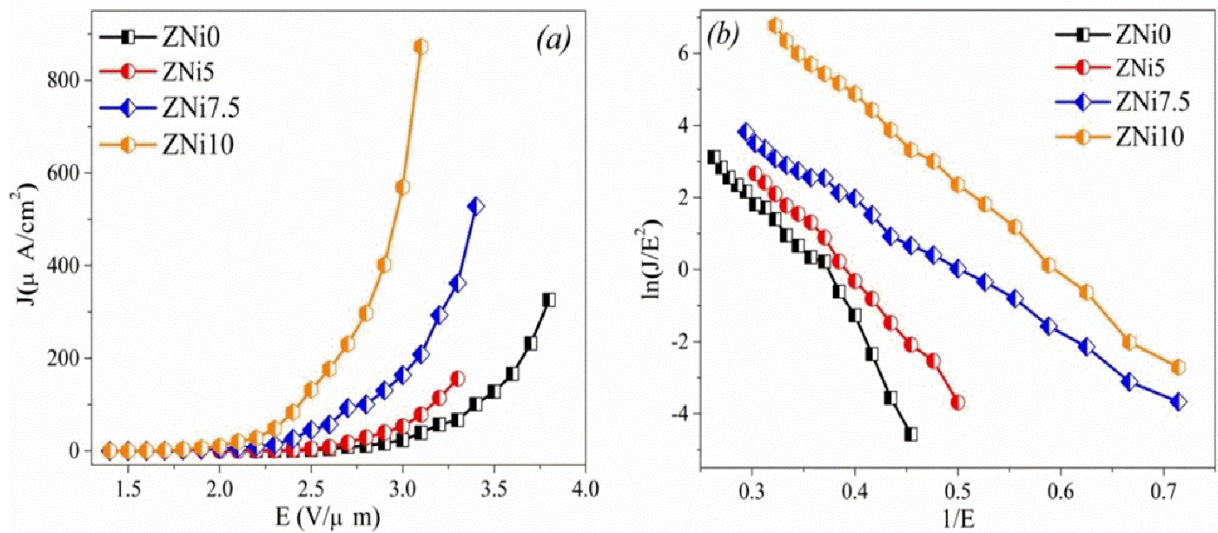

*FIG.3. Shows Field emission of pure and Ni doped ZnO. (a) Applied electrical field as a function of emission current density. (b) F-N plot.*

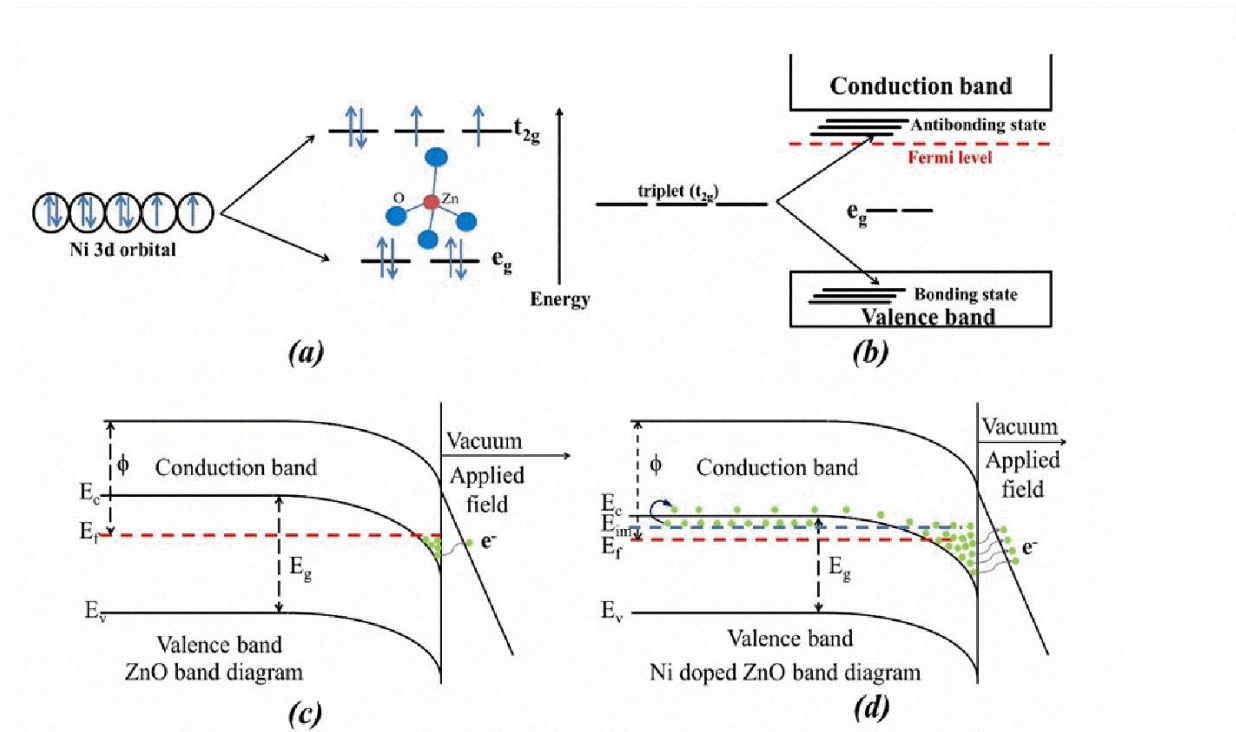

*FIG.4. Shows the schematic (a) Electronic structure of transition metal at a substitution site in a wurtzite structure. (b) Splitting of impurity state under the influence of crystal field of host ZnO. Schematic energy band diagram of (c) ZnO (d) Ni doped ZnO ($E_g$: Energy gap, $E_f$: Fermi level, $E_c$: Energy of Conduction band, $E_v$: Energy of Valance band, $\phi$: work function).*

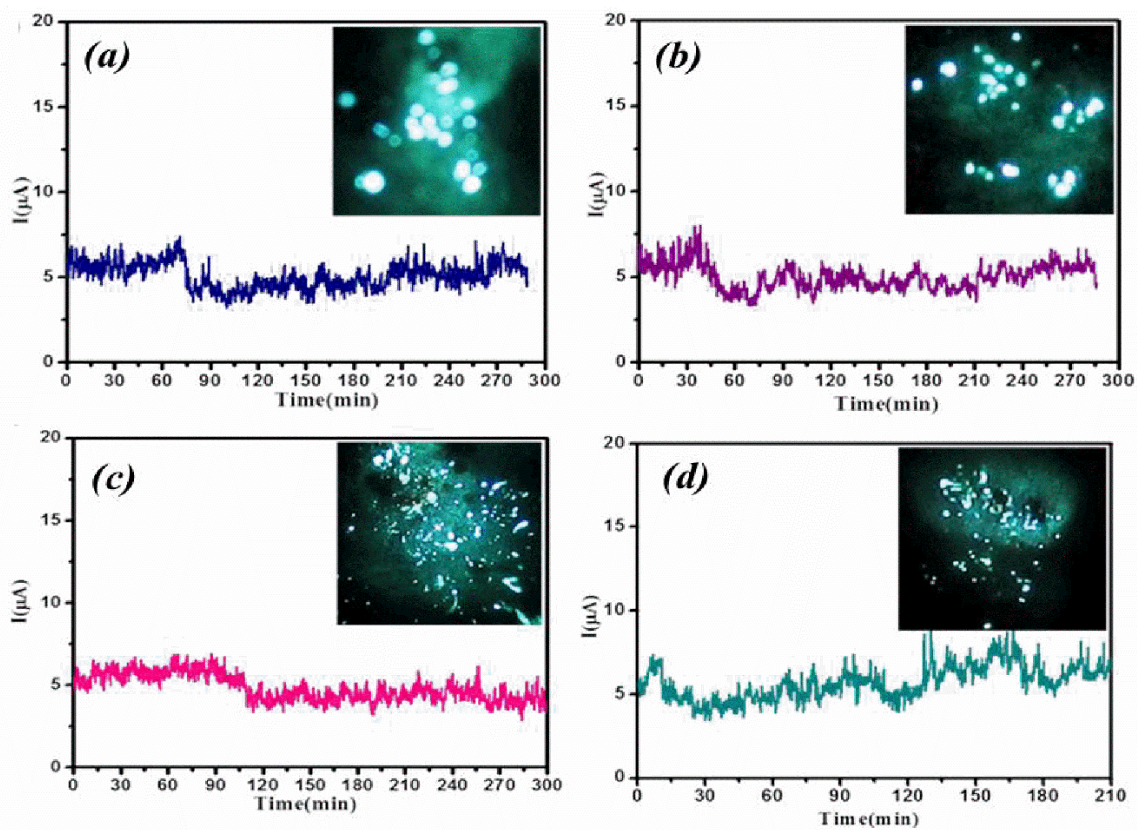

*FIG.5. Shows typical field emission current stability recorded at 5 µA indicating stable field emission current of (a) pure ZnO, (b) Ni 5% doped ZnO, (c) Ni 7.5% doped ZnO, (d) Ni 10% doped ZnO (inset of fig. shows field emission pattern taken during the long term current stability measurements of the emitter).*